\documentclass[preprint]{aastex62}
\usepackage{color}
\usepackage{times}
\usepackage{hyperref}
\usepackage{xspace}
\usepackage{amsmath}
\hypersetup{
    pdftitle={Standardization on the Rise}, 
    pdfauthor={Brian Hayden},
    colorlinks=true,        
    linkcolor=blue,         
    citecolor=blue,         
    urlcolor=blue           
}

\shorttitle{Standardization on the Rise}
\shortauthors{Hayden et al.}

\begin{document}

\newcommand{\ang}{\AA\xspace}
\newcommand{\HST}{\textit{HST}\xspace}
\newcommand{\Hubble}{\textit{Hubble Space Telescope}\xspace}
\newcommand{\WFIRST}{\textit{WFIRST}\xspace}
\newcommand{\xor}{\ensuremath{x_1^r}\xspace}
\newcommand{\xof}{\ensuremath{x_1^f}\xspace}
\newcommand{\xone}{\ensuremath{x_1}\xspace}
\newcommand{\pkg}[1]{\texttt{#1}\xspace}
\newcommand{\sncosmo}{\pkg{sncosmo}}

\newcommand{\sigmaint}{\ensuremath{\sigma_{\mathrm{unexpl}}}\xspace} 
\newcommand{\AoAPG}{\ensuremath{\alpha/(\alpha + \gamma)}\xspace}
\newcommand{\mB}{\ensuremath{\tilde{m}_B}\xspace} 
\newcommand{\mBtrue}{\ensuremath{\tilde{m}^{\mathrm{true}}_B}\xspace}
\newcommand{\ctrue}{\ensuremath{c^{\mathrm{true}}}\xspace}
\newcommand{\xortrue}{\ensuremath{x_1^{r\; \mathrm{true}}}\xspace}
\newcommand{\xoftrue}{\ensuremath{x_1^{f\; \mathrm{true}}}\xspace}
\newcommand{\xotrue}{\ensuremath{\bar{x}_1^{\mathrm{true}}}\xspace}

\newcommand{\AoAPGNominal}{\ensuremath{0.21^{+0.10}_{-0.11}}\xspace}
\newcommand{\ANominal}{\ensuremath{0.030^{+0.016}_{-0.016}}\xspace}
\newcommand{\dAlphaNominal}{\ensuremath{0.000^{+0.000}_{0.000}}\xspace}
\newcommand{\GNominal}{\ensuremath{0.115^{+0.013}_{-0.013}}\xspace}
\newcommand{\AgtGNominal}{0.1\%\xspace}
\newcommand{\GgtASigmaNominal}{3.1\xspace}
\newcommand{\AoAPGFourD}{\ensuremath{0.04^{+0.20}_{-0.21}}\xspace}
\newcommand{\AFourD}{\ensuremath{0.005^{+0.031}_{-0.030}}\xspace}
\newcommand{\dAlphaFourD}{\ensuremath{0.000^{+0.000}_{0.000}}\xspace}
\newcommand{\GFourD}{\ensuremath{0.142^{+0.029}_{-0.026}}\xspace}
\newcommand{\AgtGFourD}{1.5\%\xspace}
\newcommand{\GgtASigmaFourD}{2.2\xspace}
\newcommand{\AoAPGSkewScale}{\ensuremath{0.02^{+0.14}_{-0.16}}\xspace}
\newcommand{\ASkewScale}{\ensuremath{0.003^{+0.022}_{-0.024}}\xspace}
\newcommand{\dAlphaSkewScale}{\ensuremath{0.000^{+0.000}_{0.000}}\xspace}
\newcommand{\GSkewScale}{\ensuremath{0.145^{+0.023}_{-0.021}}\xspace}
\newcommand{\AgtGSkewScale}{0.01\%\xspace}
\newcommand{\GgtASigmaSkewScale}{3.8\xspace}
\newcommand{\AoAPGBrokenLinear}{\ensuremath{0.18^{+0.10}_{-0.11}}\xspace}
\newcommand{\ABrokenLinear}{\ensuremath{0.026^{+0.016}_{-0.017}}\xspace}
\newcommand{\dAlphaBrokenLinear}{\ensuremath{-0.020^{+0.029}_{-0.029}}\xspace}
\newcommand{\GBrokenLinear}{\ensuremath{0.117^{+0.014}_{-0.013}}\xspace}
\newcommand{\AgtGBrokenLinear}{0.06\%\xspace}
\newcommand{\GgtASigmaBrokenLinear}{3.2\xspace}
\newcommand{\AoAPGBrokenLinearFall}{\ensuremath{0.20^{+0.10}_{-0.11}}\xspace}
\newcommand{\ABrokenLinearFall}{\ensuremath{0.028^{+0.016}_{-0.016}}\xspace}
\newcommand{\dAlphaBrokenLinearFall}{\ensuremath{-0.040^{+0.031}_{-0.032}}\xspace}
\newcommand{\GBrokenLinearFall}{\ensuremath{0.113^{+0.013}_{-0.013}}\xspace}
\newcommand{\AgtGBrokenLinearFall}{0.08\%\xspace}
\newcommand{\GgtASigmaBrokenLinearFall}{3.2\xspace}
\newcommand{\AoAPGGaussPop}{\ensuremath{0.21^{+0.10}_{-0.11}}\xspace}
\newcommand{\AGaussPop}{\ensuremath{0.030^{+0.016}_{-0.016}}\xspace}
\newcommand{\dAlphaGaussPop}{\ensuremath{0.000^{+0.000}_{0.000}}\xspace}
\newcommand{\GGaussPop}{\ensuremath{0.115^{+0.013}_{-0.013}}\xspace}
\newcommand{\AgtGGaussPop}{0.09\%\xspace}
\newcommand{\GgtASigmaGaussPop}{3.1\xspace}
\newcommand{\AoAPGNominalSNRgtOne}{\ensuremath{0.19^{+0.10}_{-0.11}}\xspace}
\newcommand{\ANominalSNRgtOne}{\ensuremath{0.028^{+0.017}_{-0.017}}\xspace}
\newcommand{\dAlphaNominalSNRgtOne}{\ensuremath{0.000^{+0.000}_{0.000}}\xspace}
\newcommand{\GNominalSNRgtOne}{\ensuremath{0.121^{+0.014}_{-0.014}}\xspace}
\newcommand{\AgtGNominalSNRgtOne}{0.07\%\xspace}
\newcommand{\GgtASigmaNominalSNRgtOne}{3.2\xspace}
\newcommand{\AoAPGNominalSNRgtHalf}{\ensuremath{0.24^{+0.10}_{-0.10}}\xspace}
\newcommand{\ANominalSNRgtHalf}{\ensuremath{0.034^{+0.015}_{-0.015}}\xspace}
\newcommand{\dAlphaNominalSNRgtHalf}{\ensuremath{0.000^{+0.000}_{0.000}}\xspace}
\newcommand{\GNominalSNRgtHalf}{\ensuremath{0.109^{+0.013}_{-0.013}}\xspace}
\newcommand{\AgtGNominalSNRgtHalf}{0.3\%\xspace}
\newcommand{\GgtASigmaNominalSNRgtHalf}{2.8\xspace}
\newcommand{\AoAPGLowz}{\ensuremath{0.02^{+0.22}_{-0.26}}\xspace}
\newcommand{\ALowz}{\ensuremath{0.002^{+0.028}_{-0.029}}\xspace}
\newcommand{\dAlphaLowz}{\ensuremath{0.000^{+0.000}_{0.000}}\xspace}
\newcommand{\GLowz}{\ensuremath{0.118^{+0.022}_{-0.021}}\xspace}
\newcommand{\AgtGLowz}{0.7\%\xspace}
\newcommand{\GgtASigmaLowz}{2.4\xspace}
\newcommand{\AoAPGSDSS}{\ensuremath{0.05^{+0.15}_{-0.18}}\xspace}
\newcommand{\ASDSS}{\ensuremath{0.007^{+0.025}_{-0.026}}\xspace}
\newcommand{\dAlphaSDSS}{\ensuremath{0.000^{+0.000}_{0.000}}\xspace}
\newcommand{\GSDSS}{\ensuremath{0.145^{+0.023}_{-0.022}}\xspace}
\newcommand{\AgtGSDSS}{0.08\%\xspace}
\newcommand{\GgtASigmaSDSS}{3.1\xspace}
\newcommand{\AoAPGSNLS}{\ensuremath{0.53^{+0.18}_{-0.22}}\xspace}
\newcommand{\ASNLS}{\ensuremath{0.081^{+0.034}_{-0.036}}\xspace}
\newcommand{\dAlphaSNLS}{\ensuremath{0.000^{+0.000}_{0.000}}\xspace}
\newcommand{\GSNLS}{\ensuremath{0.070^{+0.029}_{-0.026}}\xspace}
\newcommand{\AgtGSNLS}{57\%\xspace}
\newcommand{\GgtASigmaSNLS}{-0.2\xspace}
\newcommand{\alphaTwoXOne}{\ensuremath{0.030^{+0.016}_{-0.016}}\xspace}
\newcommand{\gammaTwoXOne}{\ensuremath{0.115^{+0.013}_{-0.013}}\xspace}
\newcommand{\betaTwoXOne}{\ensuremath{3.22^{+0.13}_{-0.13}}\xspace}
\newcommand{\dbetaTwoXOne}{\ensuremath{0.74^{+0.40}_{-0.40}}\xspace}
\newcommand{\deltaTwoXOne}{\ensuremath{-0.046^{+0.021}_{-0.021}}\xspace}
\newcommand{\sigmaintLowzTwoXOne}{\ensuremath{0.102^{+0.017}_{-0.015}}\xspace}
\newcommand{\sigmaintSDSSTwoXOne}{\ensuremath{0.086^{+0.012}_{-0.012}}\xspace}
\newcommand{\sigmaintSNLSTwoXOne}{\ensuremath{0.076^{+0.015}_{-0.014}}\xspace}
\newcommand{\alphaOneXOne}{\ensuremath{0.150^{+0.009}_{-0.009}}\xspace}
\newcommand{\gammaOneXOne}{\ensuremath{0.000^{+0.000}_{0.000}}\xspace}
\newcommand{\betaOneXOne}{\ensuremath{3.07^{+0.13}_{-0.13}}\xspace}
\newcommand{\dbetaOneXOne}{\ensuremath{1.10^{+0.44}_{-0.44}}\xspace}
\newcommand{\deltaOneXOne}{\ensuremath{-0.067^{+0.021}_{-0.020}}\xspace}
\newcommand{\sigmaintLowzOneXOne}{\ensuremath{0.120^{+0.017}_{-0.016}}\xspace}
\newcommand{\sigmaintSDSSOneXOne}{\ensuremath{0.103^{+0.011}_{-0.010}}\xspace}
\newcommand{\sigmaintSNLSOneXOne}{\ensuremath{0.081^{+0.014}_{-0.013}}\xspace}

\title{SN Ia Standardization on the Rise: Evidence for the Cosmological Importance of Pre-Maximum Measurements}

\correspondingauthor{Brian Hayden}
\email{bhayden@lbl.gov}

\author{B. Hayden}
\affiliation{Department of Physics, University of California, Berkeley, CA 94720}
\affiliation{E.O. Lawrence Berkeley National Laboratory, 1 Cyclotron Rd., Berkeley, CA, 94720}
\affiliation{Space Telescope Science Institute, 3700 San Martin Drive, Baltimore, MD 21218}

\author{D. Rubin}
\affiliation{Space Telescope Science Institute, 3700 San Martin Drive, Baltimore, MD 21218}
\affiliation{E.O. Lawrence Berkeley National Laboratory, 1 Cyclotron Rd., Berkeley, CA, 94720}

\author{M. Strovink}
\affiliation{Department of Physics, University of California, Berkeley, CA 94720}
\affiliation{E.O. Lawrence Berkeley National Laboratory, 1 Cyclotron Rd., Berkeley, CA, 94720}

\begin{abstract}

We present SALT2X, an extension of the SALT2 model for SN Ia supernova light curves. SALT2X separates the light-curve-shape parameter $x_1$ into an \xor and \xof for the rise and fall portions of the light curve. Using the Joint Lightcurve Analysis (JLA) SN sample, we assess the importance of the rising and falling portions of the light curve for cosmological standardization using a modified version of the Unified Nonlinear Inference for Type Ia cosmologY (UNITY) framework. We find strong evidence that \xor has a stronger correlation with peak magnitude than \xof. We see evidence that standardizing on the rise affects the color standardization relation, and reduces the size of the host-galaxy standardization and the unexplained (``intrinsic'') luminosity dispersion. Since SNe Ia generally rise more quickly than they decline, a faster observing cadence in future surveys will be necessary to maximize the gain from this work, and to continue to explore the impacts of decoupling the rising and falling portions of SN Ia light curves. 

\end{abstract}


\section{Introduction}
Type Ia supernovae (SNe Ia) have played a key role in our understanding of the energy density of the universe, acting as ``standardizable candles'' for measuring distances and inferring the dynamics of the expansion history. They demonstrated the first strong evidence for the presence of an accelerated expansion rate \citep{riess98,perlmutter99}, and continue to provide constraints on the physics driving the acceleration \citep{scolnic18}. As the numbers of SNe used in cosmological analyses grow well into the thousands, and other sources of uncertainties (such as photometric calibration) are reduced, an improved understanding of standardization will become increasingly important for reducing the remaining uncertainties.

The nature of SN Ia standardization has been determined empirically, and historically has included three main components. 1) The ``color'' of each supernova, measured slightly differently by different light-fitting methods, is correlated with peak luminosity, likely due to a combination of dust \citep{phillips13} and an intrinsic color distribution, both requiring that bluer supernovae are brighter \citep{wang06, rubin15, mandel17}. The range of color standardizations has an RMS around $\sim 0.3$ mag. 2) The width of the light curve is positively correlated with the peak luminosity, likely due to a relationship between total radioactive energy available (the amount of \(^{56}\)Ni produced in the thermonuclear runaway of the white dwarf), and the rate of escape of optical photons in the ejecta \citep{hoeflich96, kasen07}. The light-curve-width standardization has an RMS of $\sim 0.14$ mag. 3) The final piece of the current standardization is a correlation between peak luminosity and the properties of the host galaxy; \citet{kelly10} found that supernovae in higher stellar mass host galaxies were brighter than expected after standardization, a phenomenon that has become known as the ``host mass step.'' There is increasing evidence that the host mass step is mostly driven by the age of the progenitor system \citep{rigault13,childress14,kelly15,rigault18, kim18}. The equivalent RMS of the mass step standardization is $\sim 0.05$ mag.

Given the important SN-standardization role played by light-curve width, here we focus on how that width is measured. In a standard approach \citep{jha07, guy07}, observations of a single SN Ia are fit to a family of light-curve templates in which a single width parameter controls the variation of both the rising part and the falling part of the light curve (e.g.~the ``rise time'' and ``decline rate'', suitably defined). Unfortunately for this standard approach, it is now well established that, for any fixed decline rate, the SN Ia rise time varies significantly \citep{strovink07, hayden10,ganeshalingam11}. 

In \citet{hayden10}, the ``2stretch'' model for light curve fitting was presented. In that analysis, the Sloan Digital Sky Survey (SDSS)-II \citep{frieman08} SN Ia light curves were \textit{K}-corrected to rest-frame $B$ and $V$ band, and then fit with an MLCS2k2 \citep{jha07} $\Delta=0$ template in each filter. The stretch parameter, a multiplicative factor applied to the time-axis of the light curve to estimate the width, was separated into a different stretch for the rising and falling portions of the light curve. In this work, we improve on the 2stretch model with ``SALT2X''. This is an extension of the Spectral Adaptive Lightcurve Template, version 2.4 (SALT2-4) \citep{guy07, mosher14}. We use the SALT2.4 spectral time-series surfaces but apply a different \xone to the rising (\xor) and falling (\xof) portions of the light curve. The model is described in more detail in Section~\ref{sec:salt2x}. The SALT2X model allows us to apply the premise of 2stretch more generally to a larger SN sample, leveraging the power of the SALT2 spectral template, avoiding the need for $K$-corrections, and better utilizing all photometry for each SN. The SALT2X model will be available as a ``source'' in future releases of \sncosmo \citep{sncosmo},\footnote{http://sncosmo.readthedocs.io/en/v1.5.x/} and the code to reproduce this analysis is available on GitHub.\footnote{Link to public repo will be made available upon acceptance}

Future large cadenced surveys, such as the Large Synoptic Survey Telescope \citep{lsst09} and the {\it Wide Field Infrared Survey Telescope} \citep{spergel15}, will measure thousands to tens of thousands of SNe Ia for cosmological parameter estimation. Since SNe Ia rise faster than they decline (standard practice is to include observations in the light-curve fit within -15 to 45 rest-frame days of time of maximum), accurate constraints on the rising portion of the light curve require a fast observing cadence ($\lesssim$ 4-5 rest-frame days). It is therefore critical to understand whether the rising portion of the light curves carries additional standardization information, which may help to reduce systematic uncertainties when the number of cosmologically useful SNe will grow by orders of magnitude.

In this analysis, we apply SALT2X to the Joint Lightcurve Analysis (JLA) sample of SNe Ia \citep{betoule14}. We perform a basic selection cut on the light curves, using the size and Gaussianity of the SALT2X fit posteriors as a metric for light-curve quality. We then use the Unified Nonlinear Inference for Type Ia cosmologY (UNITY) framework of \cite{rubin15} to determine the standardization parameters on the rising and falling portions of the light curve, finding a strong preference for the rising portion in the standardization. We pass a large sample of simulated light curves through the same procedure, and show that our analysis successfully recovers the input parameters.

In Section~\ref{sec:salt2x}, we present the form of the SALT2X model in terms of the standard SALT2 model. Section~\ref{sec:lcfit} describes our light-curve fits to the JLA SNe. Section~\ref{sec:dataselection} describes our data selection criteria, and Section~\ref{sec:simulation} describes our simulated data sample for testing the entire framework. In Section~\ref{sec:unity} we describe the application of the UNITY model to SALT2X, and in Section~\ref{sec:results} we present our results, including cross-checks of the analysis. We conclude and discuss the implications of our results in Section~\ref{sec:conclusions}.

\section{The SALT2X Model}
\label{sec:salt2x}

In this work, we introduce SALT2X, a version of the SALT2 light-curve model where the SALT2.4 spectral time-series surfaces are used, but separate \xor and \xof parameters are fitted, respectively, to the rising and falling portions of the light curve. Previously in \citet{hayden10}, the light curves were $K$-corrected to the Bessell $B$ and $V$ bands. SALT2X is a more extensible, accurate, and reliable procedure for adding an extra light curve width parameter to the light curve fit.

\begin{figure*}[ht]
\begin{centering}
\includegraphics[width=0.98 \textwidth]{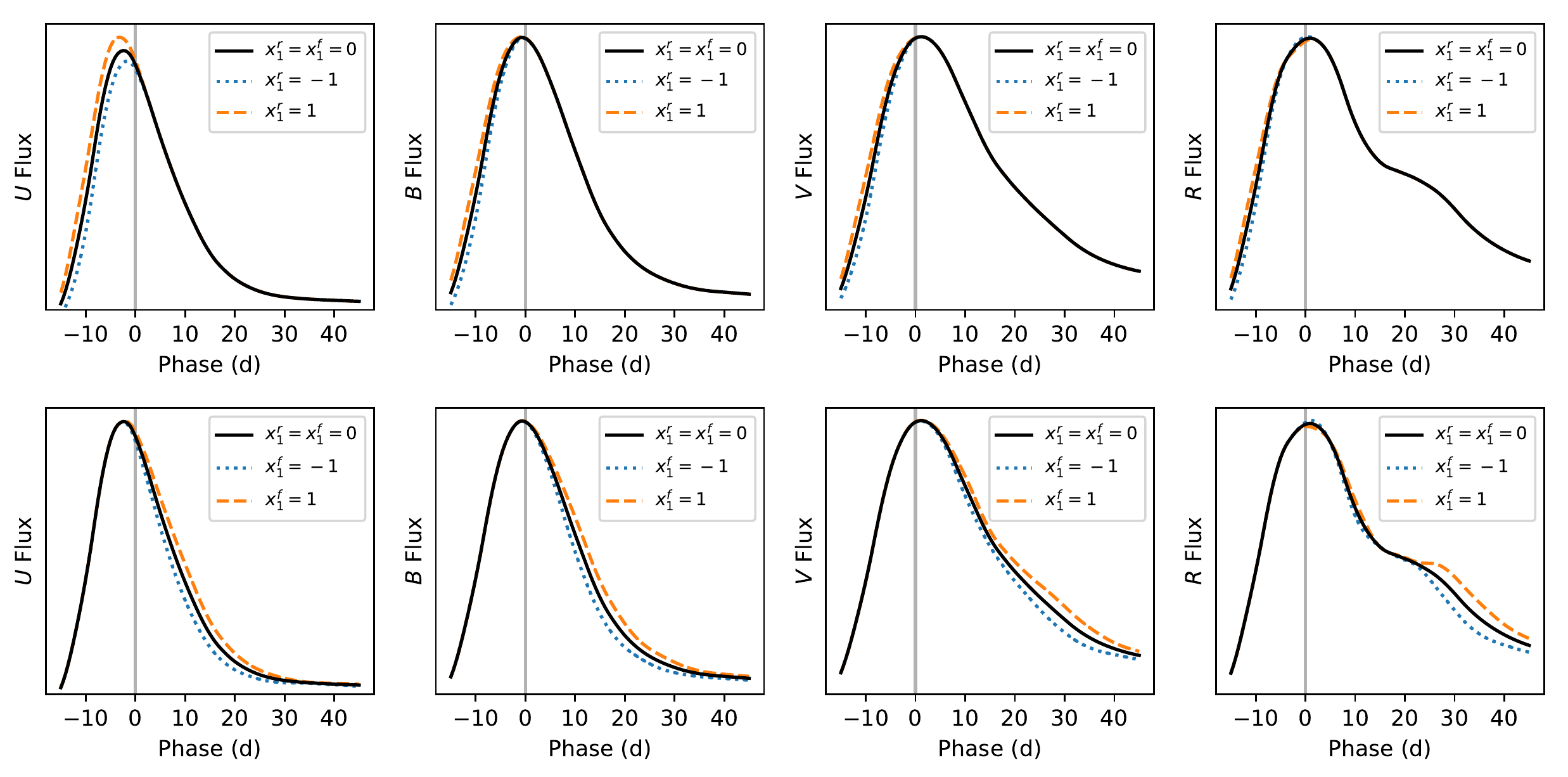}
\caption{SALT2X model light curves for rest-frame $U$, $B$, $V$, and $R$. In the top panels, we vary \xor; in the bottom, we vary \xof.}
\label{fig:lcmodel}
\end{centering}
\end{figure*}

The original SALT2 \citep{guy07} is based on the following model for the flux as a function of phase ($p$) and rest-frame wavelength ($\lambda$)
\begin{equation}
f(p,\lambda)=x_0\cdot\left[M_0(p,\lambda)+x_1\cdot M_1(p,\lambda)\right] \cdot e^{c\cdot\mathrm{CL}(\lambda)}\;,
\end{equation}
where $x_0$ is the normalization (inversely proportional to luminosity distance squared), $M_0$ is the mean model, $x_1$ is the light-curve shape parameter, $M_1$ is the variation in SED with the light-curve shape parameter, $c$ is the color parameter, and CL (the color law) is the variation (in wavelength only, not phase) with color. For the SALT2X model, we replace the single $x_1$ with a smooth function that joins \xor and \xof, matching to \xor at early phases and \xof at late phases:

\begin{equation}
x_1(p) \equiv \begin{cases}
\xor,p<-3\\
\xor + 0.5[\sin(p\frac{\pi}{6})+1](\xof - \xor),-3\le p \le 3\\
\xof,p>3
\end{cases}
\end{equation}

again, $p$ is the phase (the estimated rest-frame time of observation relative to time of maximum, $p=0$ at time of maximum). The sigmoid transition from \xor to \xof is necessary to avoid discontinuity in the light curve, since SNe Ia reach peak brightness at different times in different bandpasses. We illustrate synthesized rest-frame $U$, $B$, $V$, and $R$ light curves from our model in Figure~\ref{fig:lcmodel}.

\section{Light Curve Fitting}
\label{sec:lcfit}

Separating the rising phases of the light curve from the falling phases introduces new challenges to the light-curve fitting procedure. In particular, the JLA sample combines SNe Ia discovered in both rolling and targeted searches, so the phase coverage across surveys is not consistent. Some SNe have few observations before or after peak brightness, meaning \xor or \xof could be ill-constrained. SNe such as these will have substantially non-Gaussian uncertainties on \xor or \xof, challenging fitters that simply quote a best fit and parameter covariance matrix. We instead infer our light-curve parameters with MCMC, which properly treats non-Gaussian uncertainties. For this work we use the Python package \pkg{emcee}\footnote{http://dan.iel.fm/emcee/current/} to sample from each SN posterior. 

The SALT2X model is implemented in \sncosmo v1.5.3, using the standard SALT2.4 training. The model itself inherits from the \pkg{sncosmo.SALT2Source} class, changing only the free parameters of the model and the function for calculating the flux\footnote{We call this a \pkg{SALT2XSource}.}. This allows us to capitalize on the convenience that \sncosmo provides for many aspects of light-curve fitting, particularly filter integration, magnitude systems, and file I/O for data in the SALT2 file format.

We use filter-response curves and magnitude systems directly from the JLA data release with one exception. Since the SNLS filter response is position-dependent, and JLA does not release the filter curve for each individual SN as a unique product, we use the ``JLA-Megacam'' filters released in \pkg{SNANA}\footnote{http://snana.uchicago.edu/} to access the SN-specific filters.

We use the magnitude systems released by JLA by registering the spectral references in \pkg{sncosmo}. We apply zeropoint offsets by subtracting the zeropoints listed in Table~\ref{table:zeropoints} from the zeropoints in the JLA light-curve files. The SWOPE V-band filters are MJD-dependent as the filter was replaced in January 2006. When the filter in the JLA light-curve file is listed as ``SWOPE2::V'' the filter is set to the appropriate response curve and zeropoint via:

\[
\textrm{filter name} \equiv \begin{cases}
\textrm{swope2-v-lc3014},\textrm{MJD}<53749\\
\textrm{swope2-v-lc3009},53749\leq \textrm{MJD} \leq 53760\\
\textrm{swope2-v-lc9844},\textrm{MJD}>53760
\end{cases}
\]

Each SN has bandpasses included only if the rest-frame effective wavelength is between $3000$\AA\xspace and $7000$\AA. We use the Milky Way $E(B-V)$ reported in the JLA light-curve metadata ``MWEBV'' parameter, with the CCM89 dust model as implemented in \pkg{sncosmo v1.5.3}, applied to the model in the observer frame, and assuming $R_V=3.1$. 

The light-curve fit proceeds as follows. An initial guess for time of maximum and $x_0$ is determined by looping over a grid of dates between the earliest and latest observations of the supernova, and fitting only $x_0$ for the SALT2.4 model with $x_1=c=0$. The best $\chi^2$ point in $x_0$ and time of maximum is used to initialize the model. We then perform a full SALT2 fit using \sncosmo, which is used to cut the data to include only phases between -15 and 45 rest-frame days. This fit is then repeated once more, and another phase cut is performed at -15 to 45 rest-frame days. With this final version of the standard SALT2.4 fit, we retrieve the covariance of the SALT2X model from the SALT2.4 model covariance surfaces, and add it to the observational covariance reported by JLA in the flux covariance matrices included in the data release. The uncertainties that are used in the SALT2X fit are then fixed, and the SALT2X model covariance is no longer iterated (even though it is technically a function of \xone and $c$). This is necessary because some of the light curves with sparse rise or fall data will have \xor or \xof posteriors that span regions where the SALT2.4 model covariance is undefined. The result of this initial SALT2.4 fit is plotted, and each of these plots has been manually reviewed by eye for reasonable convergence. The pseudo log-likelihood for \pkg{emcee} is then constructed as $-0.5\times \mathbb{R}\cdot  \mathbb{C}^{-1}\cdot \mathbb{R}$, where $\mathbb{R}$ is the residual of the data and the SALT2X model, and $\mathbb{C}^{-1}$ is the inverse covariance matrix including both the SALT2X model covariance and the JLA observational covariance matrices.

With the data trimmed in phase, the model uncertainties estimated, and a log-likelihood for \pkg{emcee}, we run \pkg{emcee} with 100 ``walkers'' and 7500 samples, throwing out the first 2500 samples as burn-in. This amounts to 500,000 (100$\times$5000) samples from the posterior. For the peak apparent magnitude estimate $m_B$, used to construct the distance modulus estimate, we tried two approaches, which gave us virtually identical results in testing. The first is to make an approximate $m_B$ using $\mB \equiv -2.5 \log_{10}(x_0)$. The second is to calculate $m_B$ by constructing the best-fit SALT2X model and calculating the magnitude at peak in the Bessell $B$ filter, using the ``vega2'' JLA magnitude system. To build a posterior for $m_B$, this must be done for each MCMC chain, and becomes computationally expensive because the actual time of maximum in the $B$-band must be estimated, requiring the filter integration to be performed on a fine grid of times. 
We used $\mB \equiv -2.5 \log_{10}(x_0)$ for the results presented in this paper since it was more computationally convenient.

\section{Data Selection}
\label{sec:dataselection}

\begin{figure}[h]
\begin{centering}
\includegraphics[width=0.6 \textwidth]{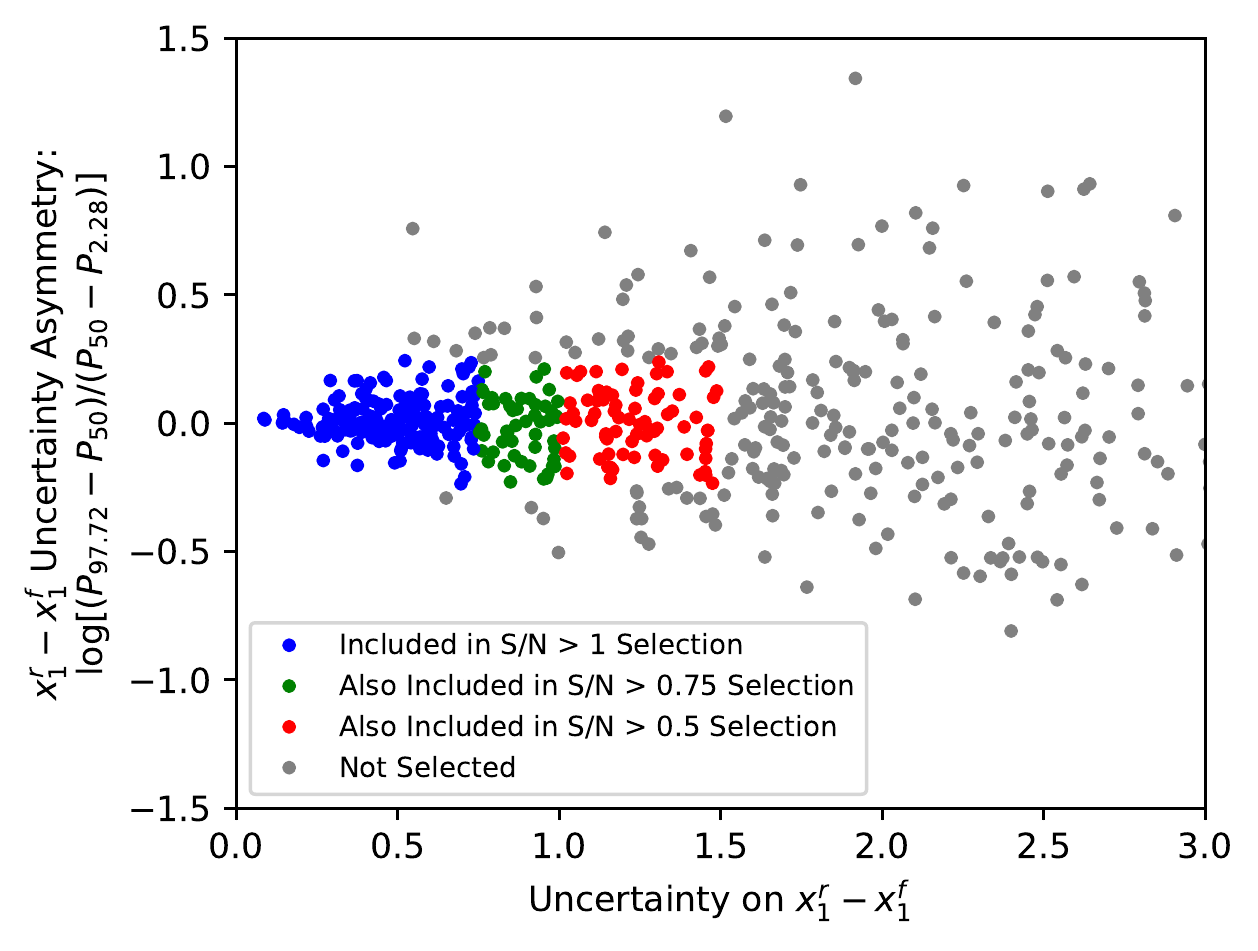}
\caption{Our percentile-based measure of the asymmetry of the $\xof - \xor$ uncertainty (Section~\ref{sec:dataselection}) plotted against the size of the uncertainty. Lower-quality light curves (the right half of the plot) have more variation in the uncertainty asymmetry. For our three sample selections, we select the SNe highlighted in blue, the blue+green (our nominal selection), and blue+green+red.}
\label{fig:dataselection}
\end{centering}
\end{figure}

As described in Section~\ref{sec:unity}, we use the Unified Nonlinear Inference for Type Ia cosmologY (UNITY) framework \citep{rubin15} to obtain our estimates of the standardization parameters. For non-outlier SNe,\footnote{UNITY uses a mixture model to simultaneously model inliers and outliers. For our analysis, we assumed that the outlier distribution has a fixed spread equal to 0.5 magnitudes in $m_B$ (added in quadrature with the other uncertainties). We do not find any SNe in our analysis where the outlier likelihood is greater than the inlier likelihood, as outliers were already rejected in building the JLA sample (their rejection was done with a frequentist analysis).} this framework assumes Gaussian light-curve fit uncertainties. However, for SNe with poorly sampled light curves, the uncertainties can be non-Gaussian, particularly for \xor or \xof. We are left with three options. 1) Compute non-Gaussian uncertainties for each SN and supply those uncertainties to UNITY (perhaps approximating these non-Gaussian uncertainties as a sum of Gaussians for computational simplicity). 2) Instead of fitting light curves as a separate, initial step, build SALT2X light-curve fits into UNITY, so that the issue of light-curve-fit parameter summary statistics is sidestepped (and thus the issue of non-Gaussian uncertainties on these parameters is sidestepped). 3) Apply a selection cut on the light-curve-fit results, selecting only well measured SNe for the analysis. As we show in Figure~\ref{fig:dataselection}, the SNe with non-Gaussian light-curve-fit uncertainties tend to be poorly measured (and thus would have much lower weight no matter our choice), so we adopt option 3), and remove these SNe from the analysis. We discuss our tests of this selection and the rest of the analysis chain in Section~\ref{sec:simulation}. These tests were performed before we saw the equivalent results for the real data. Thus, this analysis is ``blinded,'' although some of our cross-checks (Section~\ref{sec:crosschecks}) occurred to us and were performed after unblinding.

As shown in Table~\ref{tab:selectioncuts}, we perform our strongest data selection on the uncertainty on $\xof - \xor$. For our light-curve selection criteria, we define S/N to be the ability to distinguish SNe inside the distribution of a light-curve parameter. The distribution of $\xof - \xor$ has an intrinsic width of about 0.7, so S/N $>0.75$ requires $\sigma(\xof - \xor) < 1$ (see Table \ref{tab:selectioncuts} for all S/N based selection cuts and the associated uncertainty cutoffs).  We remove a few SNe with non-Gaussian (but modest) uncertainties, as shown in the remaining lines of Table~\ref{tab:selectioncuts}. Our metric for non-Gaussian uncertainties is to compare the edges of the $\sim2\sigma$ credible interval. For each of \xor, \xof, and $\xof - \xor$ we compute the 2.28$^{\rm th}$ percentile, the 50$^{\rm th}$ percentile, and the 97.72$^{\rm nd}$ percentile of the posterior samples. Then, we compute $\log{[(P_{97.72} - P_{50})/(P_{50} - P_{2.28})]}$; for a symmetric uncertainty distribution, this quantity is zero. For a skewed positive distribution, it is (almost certainly) positive, and similarly negative for negative skew.\footnote{We did not use skew directly to ensure that we considered the symmetry of only the core of the distribution and not any tails with only a small fraction of the samples.} We cut when the absolute value is larger than 0.25, indicating a significantly non-Gaussian uncertainty distribution. After selecting for modest, symmetric uncertainties, we apply a cut to remove any extreme values of \xor, \xof, or $c$, as shown in the last three lines of Table~\ref{tab:selectioncuts}. We note that these last cuts remove no SNe. All light-curve fits used in this analysis are available as an online-only table.

\begin{deluxetable*}{ccccc|c|c}[h]
\tablehead{\colhead{Selection Cut} & \colhead{SNe (Low-z)} & \colhead{SNe (SDSS)} & \colhead{SNe (SNLS)} & \colhead{Combined} & \colhead{Combined} & \colhead{Combined} \\
\colhead{} & \multicolumn{4}{c}{S/N $>$ 0.75\tablenotemark{a}} & \colhead{S/N $>$ 0.5\tablenotemark{b}} & \colhead{S/N $>$ 1\tablenotemark{c}}} 
\startdata
From JLA	&	118	&	374	&	239	&	731	&	731	&	731	\\
$\sigma(\xof - \xor) < 1/1.5/0.75$	&	62	&	105	&	72	&	239	&	349	&	177	\\
Percentile Cut $(\xof - \xor)$	&	61	&	101	&	61	&	223	&	299	&	171	\\
$\sigma (\xor) < 1.33/2/1$	&	61	&	101	&	61	&	223	&	297	&	171	\\
Percentile Cut $(\xor)$	&	61	&	95	&	57	&	213	&	274	&	165	\\
$\sigma (\xof) < 1.33/2/1$	&	61	&	95	&	57	&	213	&	274	&	165	\\
Percentile Cut $(\xof)$	&	61	&	95	&	55	&	211	&	269	&	165	\\
$-4 < \xor < 4$	&	61	&	95	&	55	&	211	&	269	&	165	\\
$-4 < \xof < 4$	&	61	&	95	&	55	&	211	&	269	&	165	\\
$-0.3 < c < 2$	&	61	&	95	&	55	&	211	&	269	&	165	\\
\enddata
\caption{Selection cuts used in our analysis. We start with the 731 Low-z + SDSS + SNLS SNe in JLA (top row), then apply sequential selection cuts and show the number of SNe remaining. The left four columns of numbers show the Low-$z$, SDSS, SNLS, and combined SNe for our nominal selection (S/N $>$ 0.75). The right two columns show other S/N cuts for just the combined sample (S/N $>$ 0.5 and 1.0). Most SNe in the sample (and most of the SNe we eliminate) are removed by our $\xof - \xor$ uncertainty cut (second row from top). The bottom three rows would remove any extreme values of \xor, \xof, or $c$, but we do not see any. The percentile cuts are the cuts on Gaussian posteriors in the light curve fit described in Section \ref{sec:dataselection}.\label{tab:selectioncuts}}
\tablenotetext{a}{S/N $>0.75$ requires $\sigma(\xof - \xor) < 1$, and both $\sigma (\xor)$ and $\sigma (\xof) < 1.33$.}
\tablenotetext{b}{S/N $>0.5$ requires $\sigma(\xof - \xor) < 1.33$, and both $\sigma (\xor)$ and $\sigma (\xof) < 2$.}
\tablenotetext{c}{S/N $>1$ requires $\sigma(\xof - \xor) < 0.75$, and both $\sigma (\xor)$ and $\sigma (\xof) < 1$.}
\end{deluxetable*}

\section{Simulated Data Generation}
\label{sec:simulation}
A test sample was constructed in order to determine how our full framework behaves for data where \xor and \xof contain equal standardization information. The goal is for the sample to have the exact phase coverage distribution as the real surveys, with known light-curve parameters and known standardization parameters. This simulated dataset provides an end-to-end test of the analysis, and imparts confidence that our results are not due to a detail of the data selection.

To accomplish this simulation, we used the real JLA epochs and uncertainties to define the observations for each simulated SN. For each JLA supernova, a SALT2X model is constructed with the redshift, time of maximum, and Milky Way $E(B-V)$ of the real supernova, with \xor, \xof, and $c$ drawn from the following covariance matrix, similar to that inferred from the real data\footnote{The simulated data were generated before the final analysis of the real data was unblinded. We noticed after the analysis was complete that the \xof/$c$ covariance should be negative. This difference in sign drives the opposite sign of the correlation between $\beta$ and $\alpha/\left(\alpha+\gamma\right)$ in Figures \ref{fig:simdata} and \ref{fig:realdata}, so despite the visual difference, we achieve end-to-end recovery of the simulation inputs.}:
\begin{align} \label{eq:covariance}
\begin{pmatrix}\xortrue\\
\xoftrue\\
\ctrue
\end{pmatrix} &\sim  N
\begin{bmatrix}
\begin{pmatrix}
0\\
0\\
0
\end{pmatrix}\!\!,&
\begin{pmatrix}
1 & 0.74 & 0\\
0.74 & 1 & 0.02\\
0 & 0.02 & 0.005
\end{pmatrix}
\end{bmatrix}
\end{align}
The absolute magnitude including standardization information is then calculated as
\begin{equation} \label{eq:simdatastandardization}
M_{B}^{\textrm{obs}}=M_{B}^{\textrm{fid}} - \gamma \cdot \xortrue - \alpha \cdot \xoftrue + \beta \cdot \ctrue + N(0,\sigmaint)
\end{equation}
where $M_{B}^{\textrm{fid}}=-19.1$, $\alpha=\gamma=0.07$, $\beta=3.1$, and $\sigmaint=0.1$. We then set this as the Bessell $B$ absolute AB magnitude of the supernova, and appropriately rescale the SALT2X $x_0$ parameter. We retrieve fluxes from the spectral time-series SALT2X model at the epochs of the JLA observations using the same bands and zeropoints as described in Section~\ref{sec:lcfit}. These fluxes are fixed to the SALT2X model, so to achieve the appropriate amount of dispersion in the photometry, we add noise drawn from a multivariate normal of the form:
\begin{equation}
\vec{n}\sim N\left[0, \mathbb{C}_{\mathrm{obs}} + \mathbb{C}_{\mathrm{model}}\right]
\end{equation}
where $\mathbb{C}_{\mathrm{obs}}$ is the covariance matrix of the measured photometric uncertainties from the JLA light curve, and $\mathbb{C}_{model}$ is the SALT2X model covariance, drawn from the SALT2.4 surfaces that describe the model uncertainty from training.\footnote{The SALT2 model covariance increases significantly at early times, so for the simulation we cap the early time model covariance to a S/N of 1, i.e., the maximum size of the model variance is (model flux)$^2$.} The simulated supernova fluxes have this noise added, and we use the flux uncertainties directly from the real JLA light curve.

A larger sample is produced by simulating twelve realizations of each JLA supernova. This simulated sample has identical phase coverage and flux uncertainties to the real light curves, but with known standardization parameters for the SALT2X model. These simulated supernovae are then run through the entire framework in the same way as the real data, including data selection, thereby testing how sensitive our results are to the cadence and uncertainties of the JLA sample. These results are discussed in Section~\ref{sec:results} and Figure~\ref{fig:simdata}. In short, we see correct recovery of the simulation inputs.

\section{UNITY and the Importance of a Bayesian Approach}
\label{sec:unity}

The initial UNITY framework was presented in \citet{rubin15}. This framework simultaneously models (nonlinear) SN standardization, cosmology fitting, the (sample-dependent) SN population, a population of outliers, systematic uncertainties, selection effects, and unexplained dispersion. Importantly, UNITY is a Bayesian hierarchical model, necessary for performing even linear regression with uncertainties in both dependent and independent variables (in this case, all the light-curve fit parameters have uncertainties), as discussed in \citet{gull89}. For each SN, latent variables describe the ``true'' values of the measurements:

\begin{align}
\begin{pmatrix}
\mB \\
\xof \\
c \\
\xor
\end{pmatrix} &\sim  N
\begin{bmatrix}
\begin{pmatrix}
\mBtrue \\
\xoftrue \\
\ctrue \\
\xortrue
\end{pmatrix}\!\!,&
C^{\mathrm{obs}} + C^{\mathrm{unexp}}
\end{bmatrix}
\end{align}

We impose the following standardization relation, which also allows us to trivially marginalize (and thus eliminate) \mBtrue:
\begin{eqnarray}
\mBtrue & = & -\ \alpha \cdot \xoftrue - \gamma \cdot \xortrue  \nonumber \\
                   &   & +\ \beta(\ctrue) \cdot \ctrue + \delta \cdot [P_{\mathrm{high}} - 0.5] + M_{i} + \mu(z_{\mathrm{helio}}, z_{\mathrm{CMB}}, \Omega_m = 0.3) \;, \label{eq:standardization}
\end{eqnarray}
where, as stated in Section~\ref{sec:lcfit}, \mB is virtually identical to the rest-frame $B$-band magnitude at peak (up to an additive normalization), but is faster to compute. Here $\alpha$ is the \xof standardization coefficient, and $\beta$ is the color standardization coefficient; as in \citet{rubin15}, we use a broken-linear color standardization, where

\begin{equation}
\beta(\ctrue) \equiv \begin{cases}
\beta_{\mathrm{Blue}} = \beta - \Delta \beta /2 , \ctrue < 0 \\
\beta_{\mathrm{Red}} = \beta + \Delta \beta /2 , \ctrue > 0 \;,
\end{cases}
\end{equation}
$\delta$ is the host-mass-standardization coefficient, and $P_{\mathrm{high}}$ is the probability that a SN host galaxy has a stellar mass $> 10^{10} M_{\odot}$. (In Section~\ref{sec:crosschecks}, we investigate a broken-linear $x_1$ standardization and find it has little effect.) $M_i$ is the estimated absolute magnitude (up to an additive constant), which we allow to be SN-sample-dependent, removing virtually all dependence of our results on the cosmological model (which we fix to flat $\Lambda$CDM with $\Omega_m = 0.3$).

As \xof is intrinsically strongly correlated with \xor, the quantity $\xof - \xor$ is generally measured only at low-to-moderate S/N, even if \xor and \xof are independently well measured. As discussed in \citet{minka99}, such low S/N regression (with uncertainties in both dependent and independent variables) must be approached with a Bayesian hierarchical model, as we do here. In such a model, informative priors are taken on \xoftrue, \xortrue, and \ctrue (representing a model of the true underlying distribution, without noise and unexplained dispersion), and the parameters in these priors (``hyperparameters'') are also included in the model. The original UNITY analysis assumed redshift- and sample-dependent Gaussian distributions for $x_1^{\mathrm{true}}$, and redshift- and sample-dependent skew-normal distributions for $c$.

We make the following changes to UNITY in this work; some of these changes are improvements, but others are merely simplifications, removing features not needed for an analysis focused on standardization rather than cosmological parameters.

\begin{itemize}
\item (Improvement) We switch to the multivariate skew-normal distribution \citep{azzalini96} describing the \xor/\xof/$c$ populations. The original UNITY analysis considered only $x_1$ and $c$, and modeled their distributions as uncorrelated. We find that the \xor and \xof distributions are intrinsically strongly correlated, so this correlation must be modeled.
\item (Improvement) We add different \xor/\xof/$c$ population means for high-mass hosted and low-mass hosted SNe. As light-curve parameters (particularly light-curve width) correlate with host-galaxy environment \citep{hamuy96, sullivan06}, there will be a (small) bias on the host-mass standardization coefficient ($\delta$) if the difference in population means is not taken into account.
\item (Simplification) We remove calibration uncertainties and selection effects. These sources of systematic uncertainty have only a small covariance with the standardization coefficients \citep{betoule14}, so we can safely exclude them, and gain a computational benefit in doing so.
\item (Simplification) We remove off-diagonal unexplained dispersion terms. The \citet{rubin15} UNITY model allowed for off-diagonal terms in the unexplained-dispersion covariance matrices. In the limit of Gaussian populations and linear standardization, these terms can describe some of the SN standardization. For example, if \ctrue has a Gaussian distribution of width 0.1 magnitudes, and the color standardization coefficient ($\beta$) is 3, then this is effectively the same as an \mB/$c$ covariance of $3 \cdot 0.1^2 = 0.03$. The original UNITY framework thus contained two types of standardization: the structural model (broken-linear relations), and the implicit linear model in the off-diagonal elements of the unexplained-dispersion covariance matrix. For this work, where we want to focus on the values of the standardization coefficients, we force these off-diagonal terms to be zero.
\end{itemize}
\vspace{15pt}

\vspace{-15pt}
With the data selected, and the updates to UNITY in place, we can investigate the standardization coefficients, which we discuss in the next section. 

\section{Results}
\label{sec:results}

\begin{figure*}[ht]
\begin{centering}
\includegraphics[width=0.55 \textwidth]{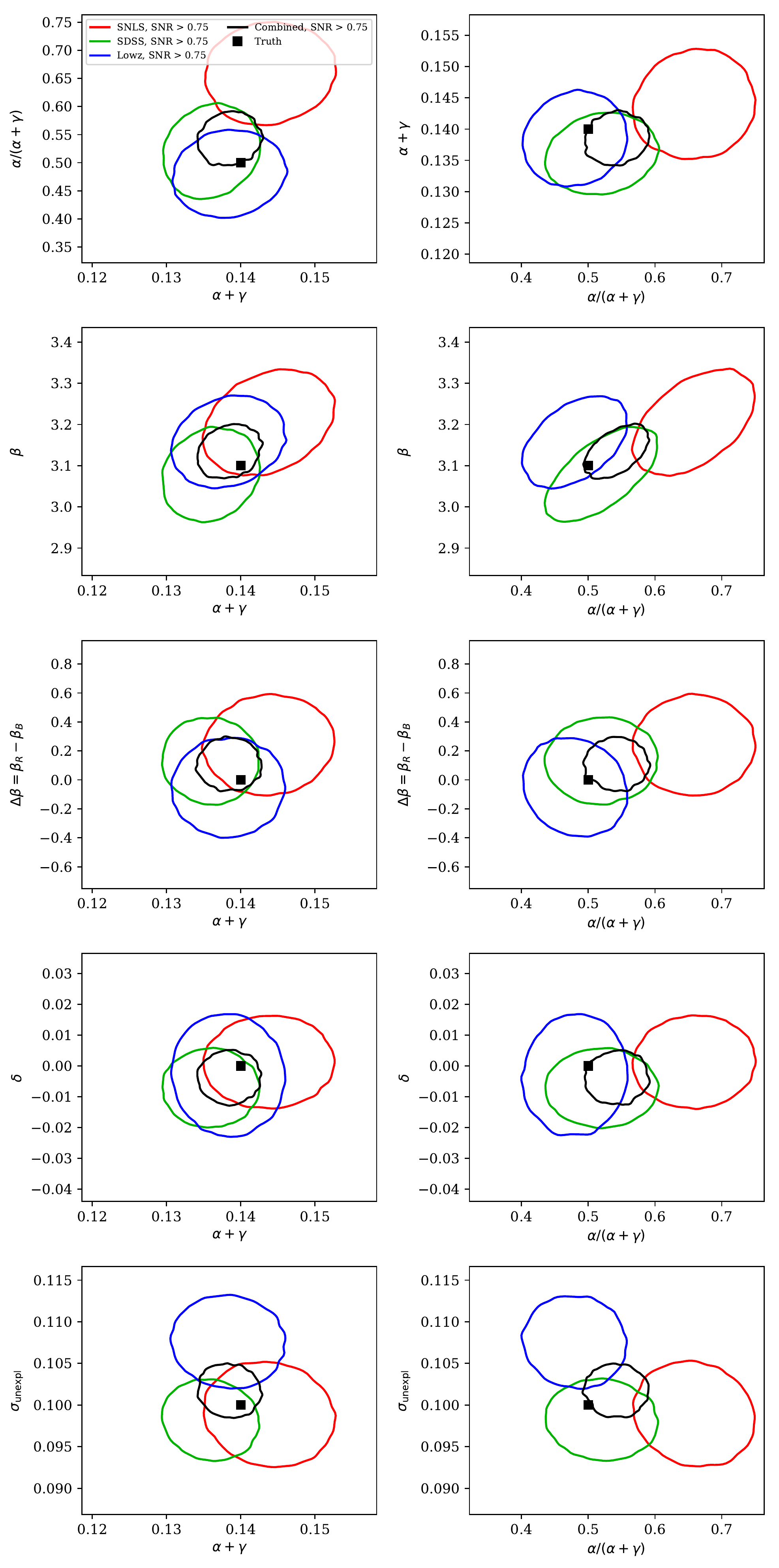}
\caption{Credible regions derived from the simulated data. Each contour is drawn based on a KDE of the MCMC samples, and encloses 68.3\% of the PDF. Low-$z$ (blue), SDSS (green), SNLS (red), and combined (black) are all shown. We mark the true simulation input coefficients with a black square. We see no evidence of biases in this dataset; in particular $\alpha$ (the \xof standardization coefficient) and $\gamma$ (the \xor standardization coefficient) are correctly recovered.}
\label{fig:simdata}
\end{centering}
\end{figure*}

We start with our recovery of the input results in the simulated data, shown in Figure~\ref{fig:simdata}. The Low-z, SDSS, SNLS, and combined constraints are shown in blue, green, red, and black, respectively. We mark the input parameters with a black square. Even with 12x the statistics of the real data, there is no evidence of biases. We also performed a simulation with 4x JLA statistics where we added covarying unexplained dispersion in both color and magnitude. We used the following values, similar to those described by \citet{kessler13} (based on \citealt{chotard11}): $C_{m_B m_B} = 9 \times 10^{-4}$, $C_{cc} = 6\times 10^{-3}$, $C_{m_B c} = 6\times 10^{-4}$, and again find no evidence of biases that affect the significance of our result. We also check the light-curve fit results against the true simulation values, and find accurate uncertainties and no evidence of biases, demonstrating end-to-end recovery from the light curve fits to the assumed standardization relation.

\begin{figure*}[ht]
\begin{centering}
\includegraphics[width=0.55 \textwidth]{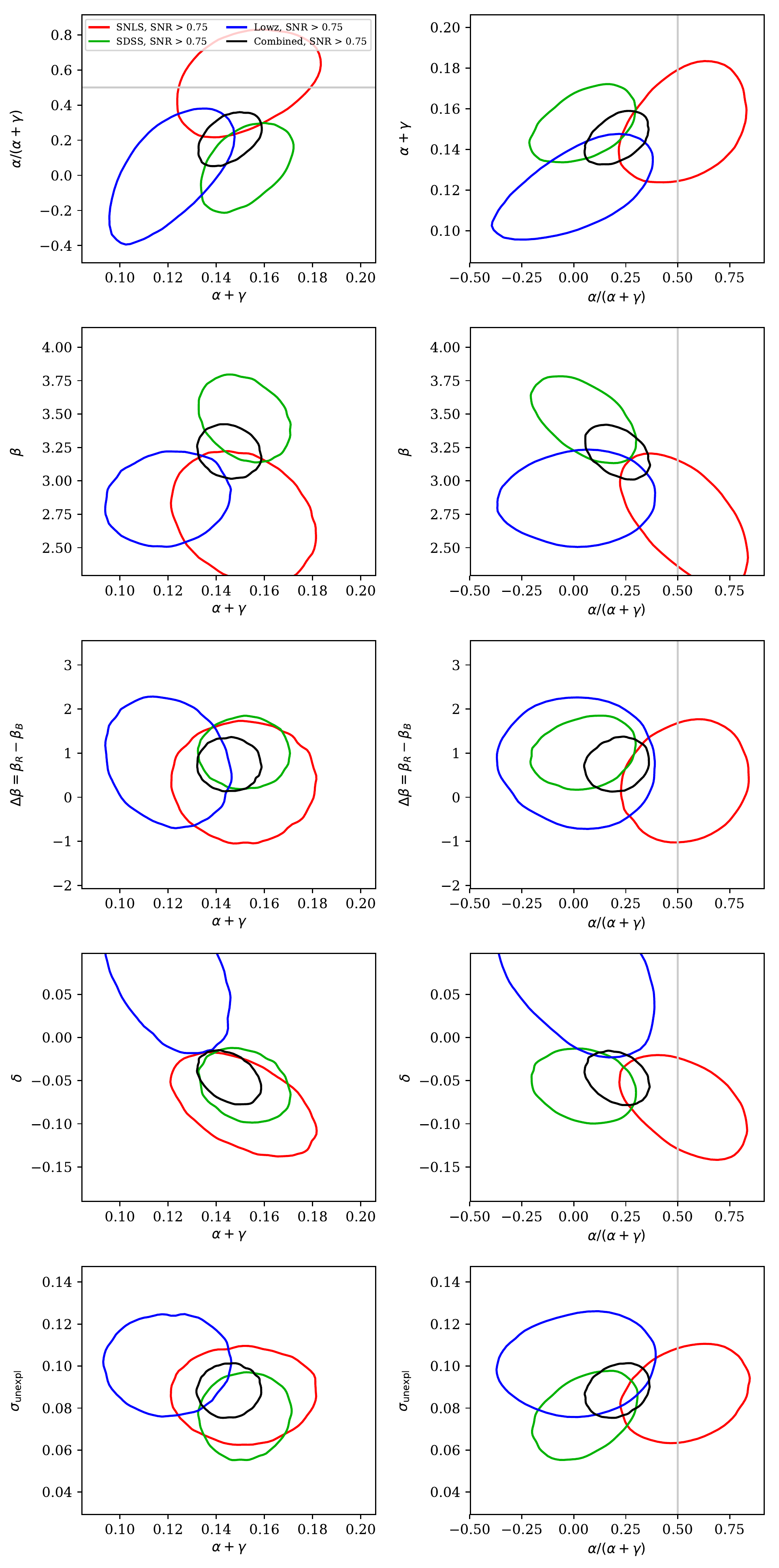}
\caption{As in Figure~\ref{fig:simdata}, we show contours enclosing 68.3\% (shaded) of the posteriors, for Low-$z$ (blue), SDSS (green), SNLS (red), and combined (black). Unlike the simulated data (Figure~\ref{fig:simdata}), there is a statistical preference for $\gamma > \alpha$, i.e., the rise-time containing more luminosity information than the decline. We also see evidence for correlations between smaller $\alpha / (\alpha + \gamma)$ and larger $\beta$, less-negative $\delta$, and smaller $\sigmaint$. For the purposes of making the combined constraints, we present the mean of all three \sigmaint values (one for each sample), rather than plotting six contours.}
\label{fig:realdata}
\end{centering}
\end{figure*}

We show similar plots for the real data in Figure~\ref{fig:realdata}, with the 68.3\% credible intervals in Table~\ref{tab:compareoneandtwo}. Unlike the simulated data (which were generated with $\alpha = \gamma$), the $\alpha/(\alpha + \gamma)$ credible interval (enclosing 68.3\% of the posterior) is \AoAPGNominal, showing a statistical preference that rise time is more important than decline time in standardization ($\gamma > \alpha$). The normalized median absolute deviation of the magnitude standardizations for the SNe in the $\textrm{S/N}>0.75$ selection cut are $\beta\cdot c=0.26$ mag, $\gamma\cdot\xor=0.13$ mag, and $\alpha\cdot\xof=0.04$ mag. The larger magnitude standardization range for \xor indicates that it is not simply a rescaled version of \xof. In the lower panels of Figure~\ref{fig:realdata}, we see that other parameters correlate with decreasing $\alpha/(\alpha + \gamma)$: $\beta$ increases, $\delta$ moves towards zero, and \sigmaint decreases. We present a comparison of credible intervals between an $\xor+\xof$ run and a single-$x_1$ run in Table~\ref{tab:compareoneandtwo}. To make this comparison fair, we use the same SNe selected for the $\xor+\xof$ run for the single-$x_1$ run. As the credible intervals are derived with the same data, the uncertainties correlate and thus the differences are generally significant. For example, going from low to high host mass moves the mean \xof by $-0.94 \pm 0.16$, while the mean of \xor moves $-0.57 \pm 0.21$. Only for 1.25\% of the MCMC samples does \xor move more than \xof. Thus, the change in $\delta$ (which is 1$\sigma$ ignoring the correlated uncertainties) is $\sim 2.2 \sigma$ taking them into account. Similarly, the correlation between $\xof - \xor$ and $c$, which drives the correlation between $\alpha/\left(\alpha+\gamma\right)$ and $\beta$, is 2.9$\sigma$.

\begin{figure}[ht]
\begin{centering}
\includegraphics[width=0.99 \textwidth]{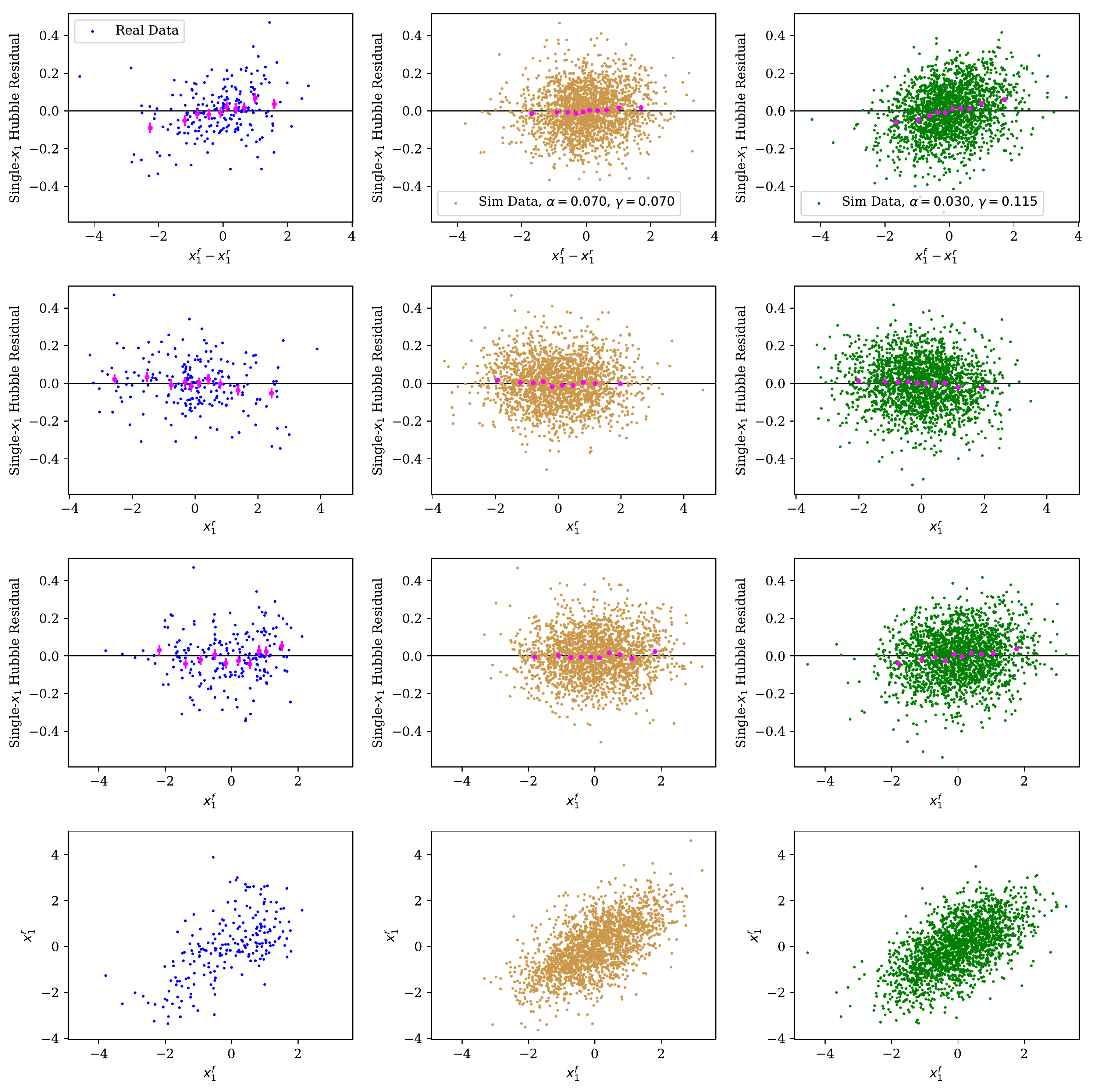}
\caption{A traditional ``Hubble residual'' view of the preference for \xor in the standardization. The left panels show results from the real data, the middle panels show simulated data with $\alpha=\gamma$, and the right panels show simulated data with $\alpha$ and $\gamma$ as observed. Top panels: Hubble diagram residuals from a single-$x_1$ analysis plotted against $\xof - \xor$. A moderate positive correlation can be seen in the real data and the $\alpha$/$\gamma$-as-observed simulation (we show binned values in magenta) as expected from our primary finding that \xor carries most of the luminosity information (Section~\ref{sec:results}). We also show single-$x_1$ Hubble residuals plotted against \xor (second-from-top panels) and \xof (second-from-bottom panels). Also as expected, the correlations here are much weaker for the real data and $\alpha$/$\gamma$-as-observed simulation, and no correlations are seen in the $\alpha=\gamma$ simulation. In the bottom panels, we show the observed \xor plotted against \xof. \xor and \xof are correlated; this must be an intrinsic correlation, as the uncertainties are almost always anticorrelated (uncertainty in the date of maximum shifts \xor and \xof in opposite directions).}
\label{fig:HRvsXone}
\end{centering}
\end{figure}

We show our main result visually in Figure~\ref{fig:HRvsXone}, which plots single-$x_1$-corrected Hubble residual against $\xof - \xor$, \xor, and \xof for the real data, simulated data with $\alpha=\gamma$, and simulated data with $\alpha$ and $\gamma$ as observed.\footnote{To get better statistics for the simulated data just for Figure~\ref{fig:HRvsXone}, we generate Gaussian random light-curve fit results, rather than performing another computationally expensive end-to-end simulation. We draw from Equations~\ref{eq:covariance} and \ref{eq:simdatastandardization}, then convolve with random draws from the light-curve-fit covariance matrices of the real data to get the values with noise. To generate a self-consistent set of single-$x_1$ values from these \xor/\xof simulates, we take the covariance-weighted mean of \xor and \xof.} It is helpful to understand these panels using a toy model. This toy model ignores the effects of finite scatter and correlations in the uncertainties, but does enable a simple visual check of the results. Suppose we define $\bar{x}_1 \equiv \frac{1}{2} (\xof + \xor)$ and $\Delta x_1 \equiv \frac{1}{2} (\xof - \xor)$. Then suppose that SN luminosity scales as $\xor$, but we standardize the luminosity with $\bar{x}_1$. In this case, part of the single-$x_1$-corrected Hubble residual should be positively correlated with $\Delta x_1$. This is exactly what we see in the top left panel of Figure~\ref{fig:HRvsXone}, which shows a positive correlation between single-$x_1$ Hubble residuals and $\xof - \xor$. As expected, we see much weaker correlations with \xor (left panel) and \xof (left bottom panel). As expected, in the middle column simulation where $\alpha=\gamma$, there is no residual correlation between single-$x_1$-Hubble residual and $\xof - \xor$, \xor, or \xof. In the right column, simulated with the same $\alpha$ and $\gamma$ as measured on the real data, we confirm the source of this residual correlation.

\subsection{Analysis Cross-Checks} \label{sec:crosschecks}

We also run a series of cross-checks on the analysis, summarized in Table~\ref{tab:analysisvariants}. We show the $\alpha/(\alpha + \gamma)$ credible interval, the fraction of the posterior with $\alpha > \gamma$ (as a measure of the statistical significance of our result), and the credible intervals for $\alpha$ and $\gamma$. In all cases, we have reasonable consistency with the nominal analysis.

\begin{deluxetable*}{ccc}[ht]
\tablehead{\colhead{Parameter} & \colhead{\xor and \xof} & \colhead{Single $x_1$ (same SN selection)}}
\startdata
$\alpha$ & \alphaTwoXOne & \alphaOneXOne \\
$\gamma$ & \gammaTwoXOne & \nodata \\
$\alpha/(\alpha + \gamma)$ & \AoAPGNominal & \nodata \\
$\beta$ & \betaTwoXOne & \betaOneXOne \\
$\Delta \beta$ & \dbetaTwoXOne & \dbetaOneXOne \\
$\delta$ & \deltaTwoXOne & \deltaOneXOne \\
Low-$z$ \sigmaint & \sigmaintLowzTwoXOne & \sigmaintLowzOneXOne \\
SDSS \sigmaint & \sigmaintSDSSTwoXOne & \sigmaintSDSSOneXOne \\
SNLS \sigmaint & \sigmaintSNLSTwoXOne & \sigmaintSNLSOneXOne \\
\enddata
\caption{Comparison of parameters obtained standardizing on both \xor and \xof and the traditional single-$x_1$ analysis. In the standardization where \xor and \xof are separate, we find a significant preference for $\gamma>\alpha$, indicating that \xor is more strongly correlated than \xof with peak magnitude. We see evidence that standardizing predominantly with \xor increases $\beta$, moves $\delta$ towards zero, and decreases $\sigmaint$. \label{tab:compareoneandtwo}}
\end{deluxetable*}

\begin{deluxetable*}{cccccc}[h]
\tablehead{\colhead{Run Variant} & \colhead{$\alpha/(\alpha + \gamma)$} & \colhead{$P(\alpha > \gamma)$} & \colhead{$\alpha$} & \colhead{$\Delta \alpha$} & \colhead{$\gamma$}}
\startdata
Nominal, S/N $> 0.75$ & \AoAPGNominal & \AgtGNominal & \ANominal & \nodata & \GNominal \\
S/N $> 1$ & \AoAPGNominalSNRgtOne & \AgtGNominalSNRgtOne & \ANominalSNRgtOne & \nodata & \GNominalSNRgtOne \\
S/N $> 0.5$ & \AoAPGNominalSNRgtHalf & \AgtGNominalSNRgtHalf & \ANominalSNRgtHalf & \nodata  & \GNominalSNRgtHalf \\
\hline
Four-Dimensional \sigmaint & \AoAPGFourD & \AgtGFourD & \AFourD & \nodata & \GFourD \\
Rescale \xor Uncertainties & \AoAPGSkewScale & \AgtGSkewScale & \ASkewScale & \nodata & \GSkewScale \\
\hline
Gaussian Populations & \AoAPGGaussPop & \AgtGGaussPop & \AGaussPop & \nodata & \GGaussPop \\
\hline
Broken-Linear $\alpha$, $\frac{1}{2}(\xof + \xor)$ & \AoAPGBrokenLinear & \AgtGBrokenLinear & \ABrokenLinear & \dAlphaBrokenLinear & \GBrokenLinear \\
Broken-Linear $\alpha$, \xof & \AoAPGBrokenLinearFall & \AgtGBrokenLinearFall & \ABrokenLinearFall & \dAlphaBrokenLinearFall & \GBrokenLinearFall \\
\hline
Low-$z$ & \AoAPGLowz & \AgtGLowz & \ALowz & \nodata & \GLowz \\
SDSS & \AoAPGSDSS & \AgtGSDSS & \ASDSS & \nodata & \GSDSS \\
SNLS & \AoAPGSNLS & \AgtGSNLS & \ASNLS & \nodata & \GSNLS \\
\enddata
\caption{Analysis variants and cross-checks. The variants on data and model selection provide a robust demonstration that $\gamma>\alpha$, consistently indicating a preference for \xor in the standardization. Two out of the three individual datasets also show a strong preference for $\gamma>\alpha$, while the third (SNLS) shows consistency with that conclusion. \label{tab:analysisvariants}}
\end{deluxetable*}

The top line shows our results for the primary analysis. The next two lines show our results varying the S/N cut.  The stability of these results is evidence that UNITY correctly treats the per-SN uncertainties.

The next two lines investigate the impact of our assumptions about the light-curve-fit uncertainties. First, we allow the unexplained dispersion term to have a component in each variable (\mB, \xor, \xof, $c$), rather than placing it in magnitude (\mB). We do note that SALT2X inherits the SALT2 model uncertainties, so some uncertainty is effectively placed in each light-curve parameter, even in the nominal analysis. Our other uncertainty test is a simple investigation of whether a pathology in the SALT2 model (e.g., incorrectly adding a large amount of model uncertainty to the rising portion of the light curves) may drive our results. In this test, we rescale all \xor uncertainties by a constant (scaling the covariance between \xor and the other parameters by the same constant). We take a broad log-normal prior on the scaling factor of $1 \pm 0.5$. These two uncertainty tests mirror each other; one changes the uncertainties by a quadrature sum, and the other by a constant. Neither test changes our main conclusion.

We also consider whether our skew-normal population distribution is driving the results. For the results in the next line, we replace the multivariate skew-normal population distribution with a multivariate Gaussian. Our conclusions are virtually unaffected.

Next, we consider a broken-linear $x_1$ standardization, as we already do for $c$. This cross-check tests whether a nonlinear \xor/\xof relation, combined with a nonlinear $x_1$/luminosity relation, drives our results. For this test, we transform our light-curve fits into the variables $\bar{x}_1 \equiv \frac{1}{2} (\xof + \xor)$ and $\Delta x_1 \equiv \frac{1}{2} (\xof - \xor)$. In analogy with Equation~\ref{eq:standardization}, we introduce a broken-linear standardization on $\bar{x}_1$:

\begin{eqnarray}
\mBtrue & = & -\ \alpha^{\prime} (\xotrue) \cdot \bar{x}_1 - \gamma^{\prime} \cdot \Delta x_1  \nonumber \\
                   &   & +\ \beta(\ctrue) \cdot \ctrue + \delta \cdot [P_{\mathrm{high}} - 0.5] + M_{i} + \mu(z_{\mathrm{helio}}, z_{\mathrm{CMB}}, \Omega_m = 0.3) \;, \label{eq:standardization}
\end{eqnarray}

where 

\begin{equation}
\alpha^{\prime} (\xotrue) \equiv \begin{cases}
\alpha^{\prime} - \Delta \alpha /2 , \xotrue < 0 \\
\alpha^{\prime} + \Delta \alpha /2 , \xotrue > 0 \;.
\end{cases}
\end{equation}

The new $x_1$ standardization coefficients are $\alpha^{\prime}$ and $\gamma^{\prime}$. We can relate these back to $\alpha$ and $\gamma$ as $\alpha = \frac{1}{2}(\alpha^{\prime} + \gamma^{\prime})$ and $\gamma = \frac{1}{2}(\alpha^{\prime} - \gamma^{\prime})$. These are the $\alpha$ and $\gamma$ values quoted in Table~\ref{tab:analysisvariants}; in addition, we also quote $\Delta \alpha$. We see a slightly negative $\Delta \alpha$ (as did \citealt{rubin15}), but it is not statistically significant and introducing $\Delta \alpha$ does not change our conclusion that $\gamma > \alpha$.

As an alternative broken-linear $x_1$ standardization, we try a broken-linear \xof standardization (keeping a linear standardization relation for \xor). This cross-check is is motivated by the observation that, for $\xof > 0$, the \xor/\xof correlation seems to be weaker (bottom panel of Figure~\ref{fig:HRvsXone}). It is thus at least possible that the luminosity changes non-linearly with \xof. Again, $\Delta \alpha$ is negative (but not statistically significant) and our conclusion that $\gamma > \alpha$ remains unchanged. Even with this freedom, $\xor$ contains more information. 

We also divide our results by dataset, shown in the last three lines of Table~\ref{tab:analysisvariants}. Two out of the three (Low-$z$ and SDSS) independently show strong evidence for $\gamma > \alpha$, and all three are consistent with the combined constraint. SNLS is the least consistent, although at least one out of three $\alpha/(\alpha + \gamma)$ subsample measurements would be expected to fall $\gtrsim 1.5$ $\sigma$ from the combined constraint more than 35\% of the time, so this is not unusual.

\section{Conclusions}

In this paper, we introduce the SALT2X model, which divides the SALT2 light-curve-shape parameter ($x_1$) into a rising (\xor) parameter and declining (\xof) parameter.
We fit the JLA sample of SNe with this model, selecting only SNe with reasonable S/N and Gaussian \xor and \xof uncertainties. In order to standardize with both parameters simultaneously (despite the correlations between them), we use UNITY, a Bayesian hierarchical model that we demonstrate correctly recovers such standardizations in the presence of such correlations. We find strong evidence that (\xof) contains only a fraction (\AoAPGNominal) of the $x_1$ luminosity information, justifying our decoupling of the rise and fall behavior. This result is robust to changes in the data selection, changing the assumed linearity of the standardization, and other analysis choices. End-to-end simulated data testing demonstrates that our result is not due to a subtle difference between the quality of the rising and falling epochs in JLA, or our implementation of the UNITY model.

When we shift more of the standardization to \xor, we see evidence that the host-mass standardization decreases in size, the unexplained luminosity dispersion decreases, and the color standardization shifts moderately in the expected direction of typical Milky-Way extinction ($\beta \sim R_V+1 = 4.1$). These findings could imply that standardizing with \xor reduces some of the astrophysical systematic uncertainties currently in SN cosmology. Thus, future surveys that seek to make SN cosmological measurements, such as the Large Synoptic Survey Telescope and the {\it Wide Field Infrared Survey Telescope} should consider maintaining, at a minimum, a cadence of one observation per 4-5 days in the rest frame to ensure that the rise and decline are independently constrained.

In \citet{hayden10}, it is noted that no significant Hubble residual effect is found by separating the rise and fall stretches. Since the SDSS sample in the SALT2X analysis demonstrates strong preference for $\gamma>\alpha$, with many of the SNe common to both analyses, we investigated the difference in conclusion regarding the importance of the rise. In Appendix \ref{app:hayden10}, we demonstrate that not including the off-diagonal covariance terms from the light-curve fitting in \citet{hayden10} leads to an effective $\chi^2$ prior that $\alpha = \gamma$. We note that in \citet{hayden10} $\alpha/\left(\alpha+\gamma\right)=0.42$, albeit without uncertainties; this qualitatively matches a detection of $\gamma>\alpha$ in the presence of a somewhat strong prior pushing toward $\alpha=\gamma$. In this way, our result is not inconsistent with that of \citet{hayden10}, but is a more thorough analysis.

In order to apply a rise-time-based analysis to a present cosmology result, one would need to include SNe with poor $\xof - \xor$ constraints. This could be handled by moving the light-curve fitting and model training inside UNITY. This allows the population parameters (which could vary with redshift) to be applied as priors for the SNe where the rise and fall are not independently measured. The unexplained dispersion could also be retrained at the same time. With the light curve fitting and training marginalized directly during the cosmology fit, uncertainties would be more easily characterized without the need for posterior distribution approximations. Such a model is computationally expensive, but worth exploring. Evaluating the best light-curve model, including the importance of the rise time, could be explored within a single framework.

\label{sec:conclusions}

\acknowledgements

We thank Kyle Barbary for helpful comments regarding integration of the SALT2X model with \pkg{sncosmo}. We also thank Greg Aldering and Kyle Boone for useful discussions throughout the analysis. We acknowledge support from NASA through the \WFIRST Science Investigation Team program. This research used resources of the National Energy Research Scientific Computing Center, a DOE Office of Science User Facility supported by the Office of Science of the U.S. Department of Energy under Contract No. DE-AC02-05CH11231.
This work was also partially supported by the Office of Science, Office of High Energy Physics, of the U.S. Department of Energy, under contract no. DE-AC02-05CH11231.

\begin{deluxetable}{cccc}
\tablehead{\colhead{filter} & \colhead{zeropoint} & \colhead{MagSys spectrum} & \colhead{System}}
\startdata
STANDARD-U & 9.724 & bd\textunderscore 17d4708\textunderscore stisnic\textunderscore 003 & Landolt 2007 \\
STANDARD-B & 9.907 & bd\textunderscore 17d4708\textunderscore stisnic\textunderscore 003 & Landolt 2007 \\
STANDARD-V & 9.464 & bd\textunderscore 17d4708\textunderscore stisnic\textunderscore 003 & Landolt 2007 \\
STANDARD-R & 9.166 & bd\textunderscore 17d4708\textunderscore stisnic\textunderscore 003 & Landolt 2007 \\
STANDARD-I & 8.846 & bd\textunderscore 17d4708\textunderscore stisnic\textunderscore 003 & Landolt 2007 \\
4SHOOTER2-Us & 9.724 & bd\textunderscore 17d4708\textunderscore stisnic\textunderscore 003 & Landolt 2007 \\
4SHOOTER2-B & 9.8744 & bd\textunderscore 17d4708\textunderscore stisnic\textunderscore 003 & Landolt 2007 \\
4SHOOTER2-V & 9.4789 & bd\textunderscore 17d4708\textunderscore stisnic\textunderscore 003 & Landolt 2007 \\
4SHOOTER2-R & 9.1554 & bd\textunderscore 17d4708\textunderscore stisnic\textunderscore 003 & Landolt 2007 \\
4SHOOTER2-I & 8.8506 & bd\textunderscore 17d4708\textunderscore stisnic\textunderscore 003 & Landolt 2007 \\
KEPLERCAM-Us & 9.6922      & bd\textunderscore 17d4708\textunderscore stisnic\textunderscore 003 & Landolt 2007 \\
KEPLERCAM-B & 9.8803       & bd\textunderscore 17d4708\textunderscore stisnic\textunderscore 003 & Landolt 2007 \\
KEPLERCAM-V & 9.4722       & bd\textunderscore 17d4708\textunderscore stisnic\textunderscore 003 & Landolt 2007 \\
KEPLERCAM-r & 9.3524       & bd\textunderscore 17d4708\textunderscore stisnic\textunderscore 003 & Landolt 2007 \\
KEPLERCAM-i & 9.2542       & bd\textunderscore 17d4708\textunderscore stisnic\textunderscore 003 & Landolt 2007 \\
SWOPE2-u & 10.514          & bd\textunderscore 17d4708\textunderscore stisnic\textunderscore 003 & Stritzinger 2011 \\
SWOPE2-g & 9.64406         & bd\textunderscore 17d4708\textunderscore stisnic\textunderscore 003 & Stritzinger 2011 \\
SWOPE2-r & 9.3516          & bd\textunderscore 17d4708\textunderscore stisnic\textunderscore 003 & Stritzinger 2011 \\
SWOPE2-i & 9.25            & bd\textunderscore 17d4708\textunderscore stisnic\textunderscore 003 & Stritzinger 2011 \\
SWOPE2-B & 9.876433        & bd\textunderscore 17d4708\textunderscore stisnic\textunderscore 003 & Stritzinger 2011 \\
swope2-v-lc3009 & 9.471276 & bd\textunderscore 17d4708\textunderscore stisnic\textunderscore 003 & Stritzinger 2011 \\
swope2-v-lc3014 & 9.476626 & bd\textunderscore 17d4708\textunderscore stisnic\textunderscore 003 & Stritzinger 2011 \\
swope2-v-lc9844 & 9.477482 & bd\textunderscore 17d4708\textunderscore stisnic\textunderscore 003 & Stritzinger 2011 \\
SDSS-u & 0.06791           & ab-spec.dat & Betoule 2012 \\
SDSS-g & -0.02028          & ab-spec.dat & Betoule 2012 \\
SDSS-r & -0.00493          & ab-spec.dat & Betoule 2012 \\
SDSS-i & -0.0178           & ab-spec.dat & Betoule 2012 \\
SDSS-z & -0.01015          & ab-spec.dat & Betoule 2012 \\
\enddata
\caption{Zeropoint offsets applied to the JLA light curve files for use in \sncosmo \label{table:zeropoints}}
\end{deluxetable}

\appendix
\section{Comparison of standardization relation with Hayden et al. 2010}
\label{app:hayden10} As mentioned in Section~\ref{sec:conclusions}, in \citealt{hayden10} (H10) the authors found $\alpha/\left(\alpha+\gamma\right)=0.42$, and based on the $\chi^2$ of the fit and the RMS of the residuals, determined that no significant preference for rise stretch (timescale) was detected. Since there is significant overlap with the JLA SDSS sample, the difference in conclusion in this work bears investigation.

There are many significant differences between the SALT2X analysis presented here and the H10 Hubble residual analysis (e.g., using the full light-curve information, rather than rest-frame $B$ and $V$, and the UNITY framework for standardization). Here we demonstrate how the lack of off-diagonal covariance terms from the light curve fits in the H10 Hubble residual analysis (source: B. Hayden, common author) acts as a prior pushing towards $\alpha=\gamma$.

\begin{figure}[ht]
\begin{centering}
\includegraphics[width=0.75 \textwidth]{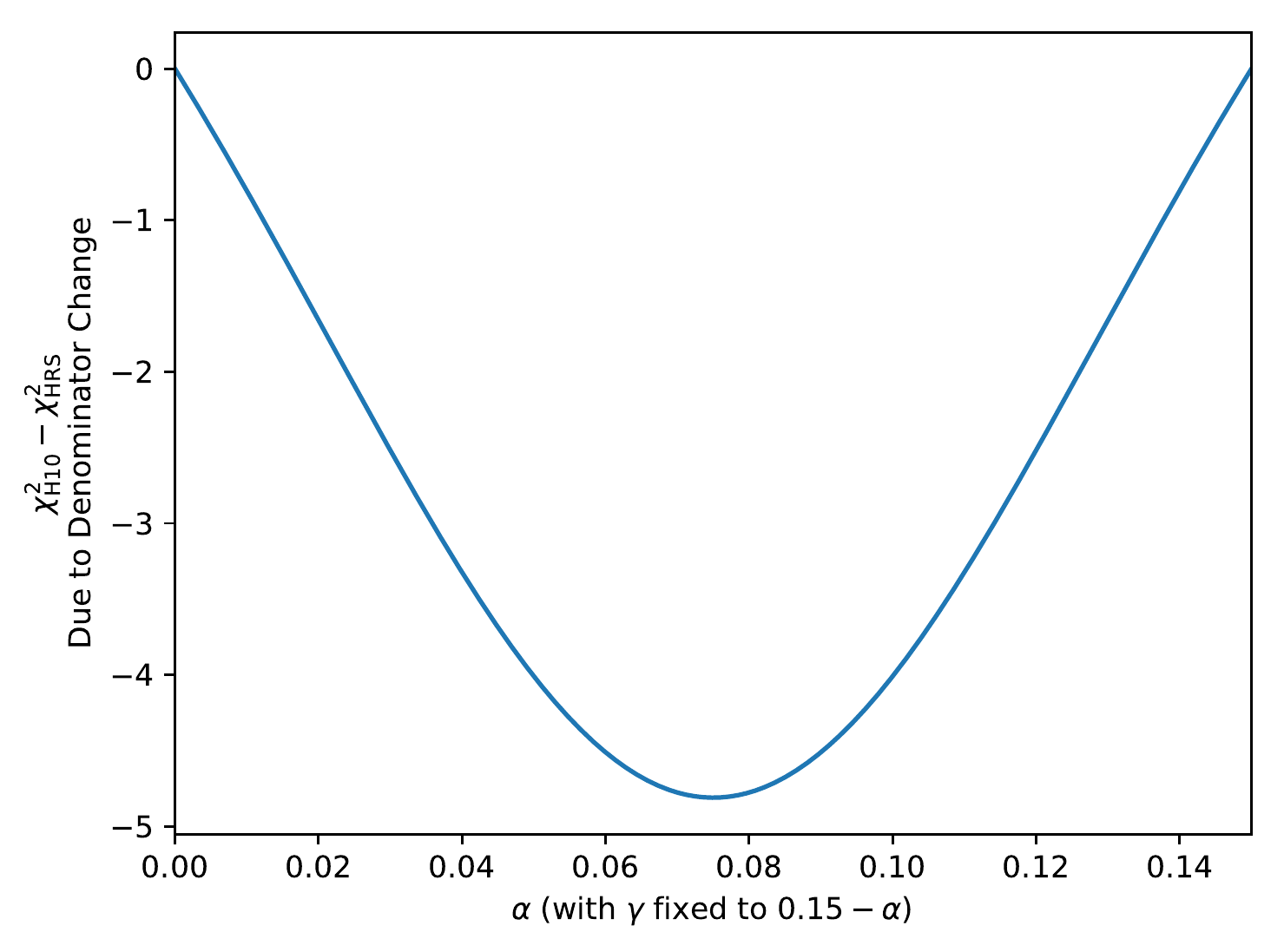}
\caption{Effective $\Delta\chi^2$ of a typical supernova due to not including the rise and fall off-diagonal covariance as in \citet{hayden10}. The lack of the covariance term behaves as a prior pushing towards $\alpha=\gamma$, where $2\,\alpha\,\gamma\,C^{\xor,\xof}$ is minimized ($C^{\xor,\xof}$ is almost always negative).}
\label{fig:delchi2}
\end{centering}
\end{figure}

We construct a pseudo-$\chi^2$ for a representative single SN both with and without the covariance between the rise and fall width measurements as follows,
\begin{align}
\chi^2_{\textrm{H10}} \propto \frac{1}{C^{\textrm{other}}+\alpha^2\,C^{\xof} + \gamma^2\,C^{\xor}} \\
\chi^2_{\textrm{HRS}} \propto \frac{1}{C^{\textrm{other}}+\alpha^2\,C^{\xof} + \gamma^2\,C^{\xor} + 2\,\alpha\,\gamma\ C^{\xor,\xof}}
\end{align}
where we use representative values for the covariance of a normal SN: $C^{\xof}=C^{\xor}=0.5$, $C^{\xor,\xof}=-0.25$, and $C^{\textrm{other}}=0.02$, which represents the combined covariance of the other terms like $\beta^2\ C^{c}$ and $C^{m_B}$. In Figure \ref{fig:delchi2}, we show $\chi^2_{\textrm{H10}}-\chi^2_{\textrm{HRS}}$ versus $\alpha$. Removal of the covariance term has a large effect, reducing the value of $\chi^2_{\textrm{H10}}$ most at $\alpha=\gamma$ where $2\,\alpha\,\gamma\,C^{\xor,\xof}$ is at an extremum. The measurement of $\alpha/\left(\alpha+\gamma\right)=0.42$ in H10 is thus a combination of the data preferring a low $\alpha$ and this prior-like $\chi^2$ difference due to the large (almost always negative) rise and fall covariance from the light-curve fits.

\end{document}